\documentclass[	
						superscriptaddress,%
						twocolumn,%
						a4paper,
						showkeys,
						endfloats
						floatfix,%
					]{revtex4-1}

\usepackage[english]{babel}
\usepackage[T1]{fontenc}
\usepackage{upgreek}
\usepackage{textcomp} 				
\usepackage{graphicx}				
\usepackage{epstopdf}				
\usepackage{color}					
\usepackage{multirow}				
\usepackage{colortbl}
\usepackage[nice]{nicefrac} 		
\usepackage{amssymb,amsmath}		
\usepackage{array}					
\usepackage{ulem}
\usepackage{float}
\usepackage{tabularx}
\usepackage{natbib}
\usepackage{longtable}

\definecolor{vert}{rgb}{0.5,0.758,0.5}
\definecolor{bleufonce}{rgb}{0,0,0.516}
\definecolor{orange}{rgb}{1,0.516,0}

\usepackage{letltxmacro}

\LetLtxMacro{\ORIGselectlanguage}{\selectlanguage}
\makeatletter
\DeclareRobustCommand{\selectlanguage}[1]{%
  \@ifundefined{alias@\string#1}
    {\ORIGselectlanguage{#1}}
    {\begingroup\edef\x{\endgroup
       \noexpand\ORIGselectlanguage{\@nameuse{alias@#1}}}\x}%
}
\newcommand{\definelanguagealias}[2]{%
  \@namedef{alias@#1}{#2}%
}
\makeatother

\definelanguagealias{en}{english}

\begin{document}

\title{Lattice dynamics and Raman spectrum of BaZrO$_3$ single crystals}
\date{\today}
\author{Constance Toulouse}
\affiliation{Materials Research and Technology Department, Luxembourg Institute of Science and Technology, 41 rue du Brill, L-4422 Belvaux, Luxembourg}
\affiliation{Physics and Materials Science Research Unit, University of Luxembourg, 41 Rue du Brill, L-4422 Belvaux, Luxembourg}
\author{Danila Amoroso}
\affiliation{Theoretical Materials Physics, Quantum Materials Center (Q-MAT), Complex and Entangled Systems from Atoms to Materials (CESAM), University of Li\`ege, Quartier Agora, All\'ee du Six Aout, B-4000 Li\`ege, Belgique}
\affiliation{Institut de Chimie de la Mati\`ere Condens\'ee de Bordeaux (ICMCB)- UMR 5026, 87, Avenue du Docteur Schweitzer, F-33608 Pessac, France}
\affiliation{National Research Council CNR-SPIN, c/o Universit\`a degli Studi ``G. D'Annunzio'', I-66100 Chieti, Italy}
\author{Cong Xin}
\affiliation{Materials Research and Technology Department, Luxembourg Institute of Science and Technology, 41 rue du Brill, L-4422 Belvaux, Luxembourg}
\affiliation{Institut de Chimie de la Mati\`ere Condens\'ee de Bordeaux (ICMCB)- UMR 5026, 87, Avenue du Docteur Schweitzer, F-33608 Pessac, France}
\author{Philippe Veber}
\affiliation{Institut Lumi\`ere-Mati\`ere (ILM) - UMR 5306, Universit\'e Claude Bernard Lyon 1, Campus LyonTech - La Doua, 10 rue Ada Byron, F-69622 Villeurbanne, France}
\affiliation{Institut de Chimie de la Mati\`ere Condens\'ee de Bordeaux (ICMCB)- UMR 5026, 87, Avenue du Docteur Schweitzer, F-33608 Pessac, France}
\author{Monica Ciomaga Hatnean}
\author{Geetha Balakrishnan}
\affiliation{Physics Department, University of Warwick, Coventry, CV4 7AL, UK}
\author{Mario Maglione}
\affiliation{Institut de Chimie de la Mati\`ere Condens\'ee de Bordeaux (ICMCB)- UMR 5026, 87, Avenue du Docteur Schweitzer, F-33608 Pessac, France}
\author{Philippe Ghosez}
\affiliation{Theoretical Materials Physics, Quantum Materials Center (Q-MAT), Complex and Entangled Systems from Atoms to Materials (CESAM), University of Li\`ege, Quartier Agora, All\'ee du Six Aout, B-4000 Li\`ege, Belgique}
\author{Jens Kreisel}
\affiliation{Physics and Materials Science Research Unit, University of Luxembourg, 41 Rue du Brill, L-4422 Belvaux, Luxembourg}
\author{Mael Guennou}
\affiliation{Materials Research and Technology Department, Luxembourg Institute of Science and Technology, 41 rue du Brill, L-4422 Belvaux, Luxembourg}
\affiliation{Physics and Materials Science Research Unit, University of Luxembourg, 41 Rue du Brill, L-4422 Belvaux, Luxembourg}

\begin{abstract}
BaZrO$_3$ is a perovskite that remains in the simple cubic phase at all temperatures, hence with no
first-order Raman-active phonon mode allowed by symmetry. Yet, it exhibits an intense Raman spectrum with sharp and well-defined features. Here, we report the evolution of the Raman spectrum of BaZrO$_3$ single crystals in a broad temperature range (4--1200~K) and discuss its origin with the support of detailed first-principle calculations of the lattice dynamics. Phonon calculations are performed not only for the cubic phase of BaZrO$_3$, but also for the low-symmetry phases with octahedra tilts that have been suspected to exist at the nanoscale. We show that the Raman spectrum shows no direct evidence for these nanodomains, but can instead be explained by classical second-order Raman scattering. We provide an assignment of the dominant features to phonon modes combinations. In particular, we show that the high frequency range of the spectrum is dominated by overtones and shows an image of the phonon density of states corresponding to the stretching modes of the oxygen octahedra.
\end{abstract}



\maketitle


\section{Introduction}

Barium zirconate (BaZrO$_3$) belongs to the family of chemically simple perovskite oxides and crystallizes in the ideal cubic structure with space group $P{m\overline{3}m}$ at ambient conditions and down to liquid helium temperatures~\cite{xin_single_2019, akbarzadeh_combined_2005}. As a functional material, BaZrO$_3$ has attracted attention as a high-temperature proton conductor \cite{kreuer_proton-conducting_2003}; it can be used in solid oxide fuel cells \cite{pergolesi_high_2010} and sensors \cite{chen_high_2009}, or as nanoparticles for increasing the vortex pinning in iron-based superconductor thin films \cite{miura_strongly_2013}. It is also of interest as one of the end-members of the (Ba,Ca)(Ti,Zr)O$_3$ system for high-performance lead-free piezoelectrics~\cite{Keeble2013_BCTZ,Liu2009,Amoroso2018,Buse2018}. At a fundamental level, its lattice dynamics and stability were found more complex than anticipated, which has drawn attention to its low-temperature structure and properties. First-principle calculations predict that the simple cubic phase is in fact slightly unstable, and that the ground state is a distorted phase of lower symmetry with octahedra tilts~\cite{akbarzadeh_combined_2005, lebedev_structural_2013, bennett_effect_2006, Amoroso2018}. Yet, a phase  transition has never been observed experimentally at ambient pressure at any temperature, but on the other hand, BaZrO$_3$ exhibits a structural phase transition under pressure around 11~GPa, presumably towards a tetragonal tilted phase \cite{chemarin_high-pressure_2000}. Such a large stability range of the cubic phase is compatible with the Goldshmidt tolerance factor of 1 in BaZrO$_3$ but is the exception rather than the rule in simple perovskite oxides. The overwhelming majority of ABO$_3$ compounds are distorted at ambient conditions and their cubic phase, when experimentally observable, is only stabilized at high pressures or temperatures: for example, above 1300~K in CaTiO$_3$~\cite{Redfern1996}, and above 820~K or 14~GPa in LaAlO$_3$~\cite{Hayward2005,Guennou2011_LAO}.

Distorted perovskites all possess first-order Raman-active modes, some of them directly related to the corresponding order parameters. In contrast, simple cubic perovskites have no Raman-active phonon mode so that no first-order Raman spectrum is allowed by symmetry. In spite of this, past studies of BaZrO$_3$ have revealed an intense Raman spectrum~\cite{xin_single_2019,Helal2016,Charrier-Cougoulic1999,Colomban2009,chemarin_high-pressure_2000} with relatively sharp features. Similar situations have been encountered in some other cubic perovksites. The most well-known is the incipient ferroelectric  SrTiO$_3$ whose Raman spectrum has been studied in detail~\cite{Fleury1968,Nilsen1968,Perry1967,Stirling1972,Grzechnik1997}. Other cases include KTaO$_3$~\cite{Nilsen1968},  BaSnO$_3$~\cite{stanislavchuk_electronic_2012}, or even BaTiO$_3$ in its cubic phase above $T_c$~\cite{Burns1978}. In general, the interpretation and understanding of Raman spectra in cubic perovskites is not straightforward, and the assignment of Raman bands to precise structural or vibrational features requires more extensive experimental and theoretical work.  

Two types of interpretations have been put forward in past studies to explain Raman spectrum of simple cubic perovksites. The first explains the spectra by second-order Raman processes, where two phonons with opposite parallel wave-vectors are involved  in the same scattering event. Second-order scattering is always allowed; it is expected to be weaker in intensity, but, in the absence of first-order scattering, it may become the dominant process. Conservation of momentum for second-order processes no longer restricts the phonons involved to the center of the Brillouin zone, and selection rules then allow for a large number of possible combinations. An accurate assignment of Raman features to specific vibrations or combination of vibrations then requires a comprehensive knowledge of the whole phonon dispersion, which may come from experiment or lattice dynamics calculations - as shown by the long history of SrTiO$_3$ second-order Raman interpretations ~\cite{Fleury1968,Nilsen1968,Perry1967,Stirling1972}. Salient features in the Raman spectrum are typically assigned to combinations of zone-boundary phonon modes where singularities in the density of states are expected.

The second interpretation invokes some local symmetry breaking that relaxes the Raman selection rules and allows for Raman scattering to occur, while averaging to a cubic symmetry at a macroscopic (XRD) scale. This is typically the case in BaTiO$_3$ where Ti-displacements remain correlated above $T_c$ in the macroscopic cubic phase, which results in the persistence of Raman peaks, but also deviations in other physical properties, like birefringence~\cite{ziebinska_birefringence_2008}. In this case, the Raman scattering process is rather of first-order in nature. This explanation has been proposed for BaZrO$_3$ also~\cite{chemarin_high-pressure_2000,Lucazeau2003,Colomban2009}, whereby the local distortions could be caused by octahedra tilts, and related to the structural instability revealed by DFT. However, the detailed interpretation remains an open problem. This has largely to do with the fact that, experimentally, BaZrO$_3$ single crystals are difficult to grow because of their high melting point. As a result, it has been usually studied as polycrystalline samples~\cite{Charrier-Cougoulic1999,Colomban2009}, and comparatively less than other perovskites.

In this work, we use recently synthesized BaZrO$_3$ single crystals~\cite{xin_single_2019} to investigate the lattice dynamics by a combination of single crystal Raman spectroscopy in a broad range of temperatures at ambient pressure, Density Functional Theory (DFT)-based calculations of the dispersion and mode symmetry analysis. We demonstrate that the Raman spectrum of BaZrO$_3$ is dominated by second order Raman scattering, without any evidence for octahedra tilts at the local scale. We also provide assignments of the most intense modes of the Raman spectra in terms of vibration pattern.

\section{Methods \& experimental details}

The single crystals used in this work have been described in Ref.~\cite{xin_single_2019}. They include a crystal grown by the floating zone technique and two crystals purchased from Crystal~Base~Co.~Ltd. and grown by the tri-arc method~\cite{reed_tri-arc_1968, xin_single_2019}. No significant difference has been observed in the Raman spectra of the different crystals. Polarized Raman spectra were recorded in a back-scattering geometry with a Renishaw inVia Reflex Raman Microscope using the 442~nm laser line of a He-Cd laser with an output power of 58~mW. Low temperature measurements have been performed using an Oxford Microstat-Hire open-cycle He cryostat together with an Oxford Mercury ITC controller. High temperature measurements have been recorded inside a TS1500 Linkam cell.

First-principles calculations of structural and dynamical properties rely on DFT and density functional perturbation theory (DFPT)\cite{dfpt}, as implemented in the ABINIT package~\cite{abinit1,abinit2}. The exchange-correlation potential was evaluated within the generalized gradient approximation (GGA) using the Wu-Cohen (WC) functional~\cite{wc-gga}. Norm-conserving pseudopotentials~\cite{pseudo} have been employed with the following orbitals considered as the valence states: $5s$, $5p$, and $6s$ for Ba $4s$, $4p$, $4d$, and $5s$ for Zr, and $2s$ and $2p$ for O. The energy cutoff for the expansion of the electronic wave functions has been fixed to 45 Ha and we used a 6$\times$6$\times$6 k-point mesh for the Brillouin zone sampling. Phonon calculations have been performed on the fully relaxed cubic and antiferrodistortive (AFD) structures, for which the structural optimization reached values of below $\sim 10^{-4}$~eV/\AA.
The q points for the phonon dispersion curves of the 5-atom cubic BZO included $\Gamma$,X, M, R, and the $\Lambda$ point halfway from $\Gamma$ to R of the simple cubic Brillouin zone.

Symmetry analysis of the phonon modes was done with the help of programs from the Bilbao crystallographic server~\cite{kroumova-bilbao-2003,elcoro-bilbao-2017} and the ISOTROPY Software Suite~\cite{isotropy}.

\section{Results}

\subsection{Lattice dynamics calculations}

\begin{figure*}[htpb]
\resizebox{15cm}{!}{\includegraphics{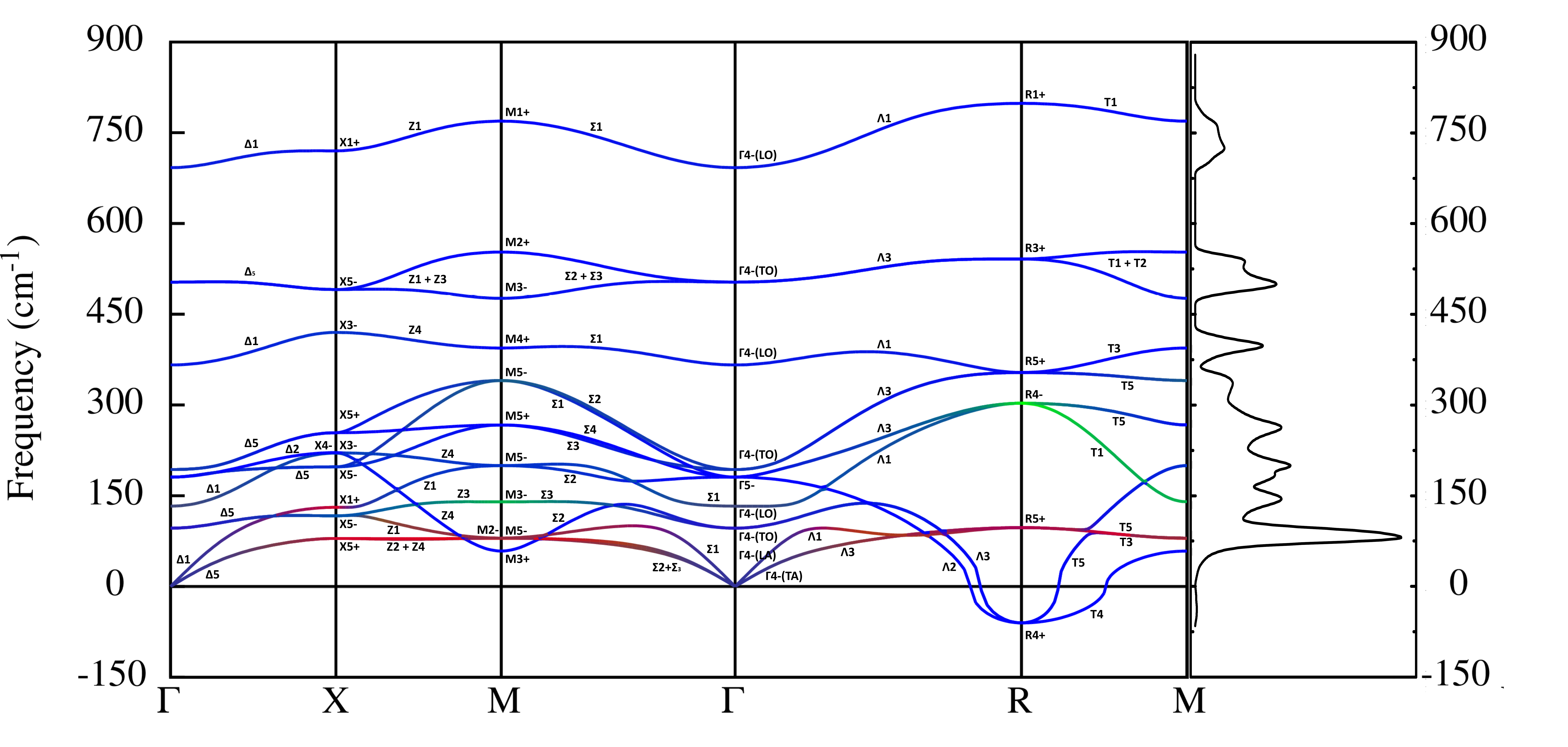}}
\caption{Dispersion curves of the phonons modes of BaZrO$_3$ in its cubic phase computed by DFPT. Negative values of frequencies refer to imaginary frequencies. Colors correspond to the atomic type involved in the vibration : blue for Oxygen, green for Zirconium and red for Barium atoms. The total density of states (DOS) is plotted on the right.}
\label{DFT_curves}
\end{figure*}

We first computed the lattice dynamics for the $Pm\bar{3}m$ cubic phase over the full Brillouin zone. The dispersion relations together with the symmetry assignment for the modes and the branches is given in Figure~\ref{DFT_curves}. This is essentially in agreement with previous reports of similar calculations~\cite{akbarzadeh_combined_2005, bennett_effect_2006}. We find that the cubic phase is dynamically unstable, with a unstable phonon mode at the $R$ point of the Brillouin zone corresponding to the tilts of the ZrO$_6$ octahedra. 

There are few experimental studies on the lattice dynamics of cubic BaZrO$_3$ that can be used for comparison. Only the polar modes have been measured by IR spectroscopy on polycristalline samples~\cite{nuzhnyy_broadband_2012}. The three polar modes in cubic perovskites are commonly referred to as ``Slater" mode (anti-phase motion of the B cation with respect to the oxygen octahedra), ``Axe" mode (anti-phase motion of the apical oxygens O$_5$ with respect to the other oxygen atoms of the octahedra) and ``Last" mode (anti-phase motion of the A cation and the ZrO$_6$ octahedra) as described in ferroelectric transitions~\cite{hlinka2006}. Their DFT-computed frequencies are in agreement with experimental IR values from ref \onlinecite{nuzhnyy_broadband_2012} within 20~cm$^{-1}$ (our calculated frequencies of the TO modes are 96, 193 and 503~cm$^{-1}$ respectively, for 116, 214 and 520~cm$^{-1}$ reported in ref \onlinecite{nuzhnyy_broadband_2012}). The LO frequencies show a better agreement, with calculated (resp. experimental) values at 133 (141), 366 (380) and 692 (692)~cm$^{-1}$.

We then calculated the lattice dynamics for the phases resulting from the condensation of the $R$-point phonon mode, namely the tetragonal $I4/mcm$, orthorhombic $Imma$ and rhombohedral $R\bar{3}c$ phases corresponding to out-of-phase oxygen rotations around one, two and three directions of the pseudo-cubic reference cell, respectively. All three AFD phases are dynamically stable, as obtained by the calculation of the phonon dispersion curves (see \textcolor{blue}{Fig. 2 in Suppl.}), and lower in energy than the cubic phase. Maximum energy gain is obtained with octahedra rotation around [001] direction in the tetragonal phase, which displays the largest distortion, as shown in \textcolor{blue}{Fig.~1 of Suppl}. Nevertheless, those phases can be considered quasi-degenerate, the difference in energy being within 0.3~meV/f.u.~\cite{Amoroso2018}.  

Details of the different phonon modes, along with their symmetry assignments, for all 4 phases and their correspondences between phases is shown on Table \ref{DFT_modes_freq}. For each phase, we identified the predicted Raman active modes and their relation to the modes in the parent phase. In all cases, the Raman-active modes originate from modes at the $R$ point folded onto the $\Gamma$ point. 
In principle, Raman spectroscopy can distinguish between these structural variants because they have different numbers of Raman modes of different degeneracies and symmetries: the tetragonal phase has 7 Raman modes (A$_{1g}$+B$_{1g}$+2B$_{2g}$+3E$_g$), the orthorhombic phase has 12 (3A$_g$+2B$_{1g}$+3B$_{2g}$+4B$_{3g}$) and the rhombohedral phase has 5 (A$_{1g}$+4E$_g$). However, in practice, we observe that these modes originate in most cases from the same degenerate vibrations of the cubic phase, and the very small distortion makes the mode splitting extremely small, sometimes below the resolution limit of classical instruments. Finally, at the level of the phonon DOS, the three phases look almost the same (supplementary material). The notable exceptions are the octahedra breathing mode, which is only Raman active in the orthorhombic phase, and the soft tilt modes, whose frequencies vary greatly depending on the tilt system. This, combined to the fact that well-oriented single domains are difficult to obtain when going through a ferroelastic transition, makes it non-trivial to identify which phase is actually stabilized. 

\begin{table*}[th!]
	\centering
	\caption{Optical phonon modes frequencies and symmetries in the different structural phases calculated by DFT (acoustic modes are omitted). The different modes are grouped together based on the symmetry transformations between phases and the modes indicated in bold font correspond to the Raman active modes (for first order processes).}
	\label{DFT_modes_freq}
	\setlength{\extrarowheight}{2pt} 
	\begin{tabularx}{\textwidth}{cc|l|cc|cc|cc}
		\hline\hline
		\multicolumn{2}{l|}{Cubic ($Pm\bar{3}m$ - 221)} & Related atomic displacements &
		\multicolumn{2}{r|}{Tetra ($I4/mcm$ - 140)} & \multicolumn{2}{r|}{Ortho ($Imma$ - 74)} & \multicolumn{2}{r}{Rhombo ($R\bar{3}c$ - 167)}\\
		\multicolumn{2}{l|}{TO/LO modes} & & \multicolumn{2}{r|}{} & \multicolumn{2}{r|}{} & \multicolumn{2}{r}{}\\
		\hline
		\multirow{3}{*}{-60} & \multirow{3}{*}{R$_4^+$} & & \textbf{23} & \textbf{Eg} & \textbf{20} & \textbf{B2g} & \textbf{73} & \textbf{Eg}\\
		 & & oxygen octahedra rotations & & & \textbf{23} & \textbf{B1g} & & \\
		 & & & \textbf{93} & \textbf{A1g} & \textbf{87} & \textbf{Ag} & \textbf{88.5} & \textbf{A1g}\\
		\hline
		\multirow{3}{*}{97} & \multirow{3}{*}{R$_5^+$} & & \textbf{102} & \textbf{Eg} & \textbf{100} & \textbf{B2g} & \textbf{104} & \textbf{Eg}\\
		 & & antiparallel Ba motion & & & \textbf{101} & \textbf{B3g} & & \\
		 & & & \textbf{104} & \textbf{B2g} & \textbf{107} & \textbf{Ag} & 98 & A2g\\
		\hline
		\multirow{3}{*}{96/133} & \multirow{3}{*}{$\Gamma_4^-$ (T1u)} & zone center ``Last" modes: & 107 & Eu & 104 & B3u & 112 & Eu\\
		 & & Ba displacement against & & & 107 & B2u & & \\
		 & & ZrO$_6$ octahedra & 113 & A2u & 114 & B1u & 99 & A2u\\
		\hline 
		\multirow{3}{*}{181} & \multirow{3}{*}{$\Gamma_5^-$ (T2u)} & silent zone-center mode: & 192 & Eu & 186 & B3u & 188 & Eu\\
		 & & antiparallel out-of-plane displace- & & & 194 & B1u & & \\
		 & & -ment of in-plane oxygens & 195 & A2u & 192 & Au & 193 & A2u\\
		\hline
		\multirow{3}{*}{193/366} & \multirow{3}{*}{$\Gamma_4^-$ (T1u)} & zone center ``Slater" modes: & 195 & Eu & 194 & B2u & 195 & Eu\\
		 & & out-of-plane displacements & & & 194 & B3u & & \\
		 & & of Zr against in-plane O & 200 & B1u & 204 & B1u & 204 & A1u\\
		\hline
		 \multirow{3}{*}{303} & \multirow{3}{*}{R$_4^-$} & \multirow{3}{*}{Zr antiparallel displacements} & 301 & A1u & 301 & Au & 301 & A1u\\
		 & & & 302 & Eu & 302 & B1u & 303 & Eu\\
		 & & & & & 302 & B2u & & \\
		\hline
		\multirow{3}{*}{354} & \multirow{3}{*}{R$_5^+$} & oxygen octahedra & \textbf{356} & \textbf{Eg} & \textbf{354} & \textbf{B2g} & \textbf{358} & \textbf{Eg}\\
		 & & shearing modes & & & \textbf{356} & \textbf{B3g} & & \\
		 & & & \textbf{358} & \textbf{B$_\mathrm{2g}$} & \textbf{360} & \textbf{A$_g$} & 353 & A2g\\
		\hline
		\multirow{3}{*}{503/692} & \multirow{3}{*}{$\Gamma_4^-$ (T$_{1u}$)} & zone center ``Axe" modes: & 494 & Eu & 494 & B3u & 497 & Eu\\
		 & & displacement of apical oxygens & & & 495 & B2u & & \\
		 & &  against in-plane oxygens & 496 & A2u & 497 & B1u & 494 & A2u\\
		\hline
		541 & R$_3^+$  & Jahn-Teller-like distorsions& 534 & A2g & \textbf{535} & \textbf{B1g} & \textbf{535} & \textbf{E$_g$}\\
		 & & of oxygen octahedra & \textbf{535} & \textbf{B$_{1g}$} & \textbf{535} & \textbf{B3g} & &\\
		\hline
		799 & R$_1^+$ & oxygen octahedra breathing mode & 791 & A$_{2g}$ & \textbf{791} & \textbf{B$_{3g}$} & 792 & A$_{2g}$\\
		\hline
		\hline\\
	\end{tabularx}
\end{table*}

In addition, to investigate the hypothesis of a second-order Raman spectrum, we determined the symmetries of the two-phonon states for all phonons at the zone center and at the high-symmetry points at the zone boundary ($\Gamma$, $R$, $X$ and $M$) where peaks and kinks in the second-order spectrum are expected due to van Hove singularities~\cite{cardona_light_1982}. For a cubic perovskite with point group $m\overline 3m$, the problem is in principle similar to the cases of silicon, diamond, or cubic SiC ~\cite{temple_multiphonon_1973,windl_second-order_1993,Windl1994}, only with many more possible phonon combinations for second-order scattering. The point group $m\overline 3m$ has three Raman-active representations A$_\mathrm{1g}$, E$_\mathrm{g}$ and T$_\mathrm{2g}$. Tables providing the symmetry of combinations for special k points and relevant phonon symmetries in BaZrO$_3$ and cubic perovskites in general are provided in the supplementary material. An important and general result is that overtones are always Raman-active. Besides, the majority of combinations contain more than one Raman-active representation, so that, generally speaking, very few of them would exhibit clear selection rules. 

\subsection{Temperature dependent Raman spectroscopy}

\begin{figure}[h!]
    \resizebox{8.6cm}{!}{\includegraphics{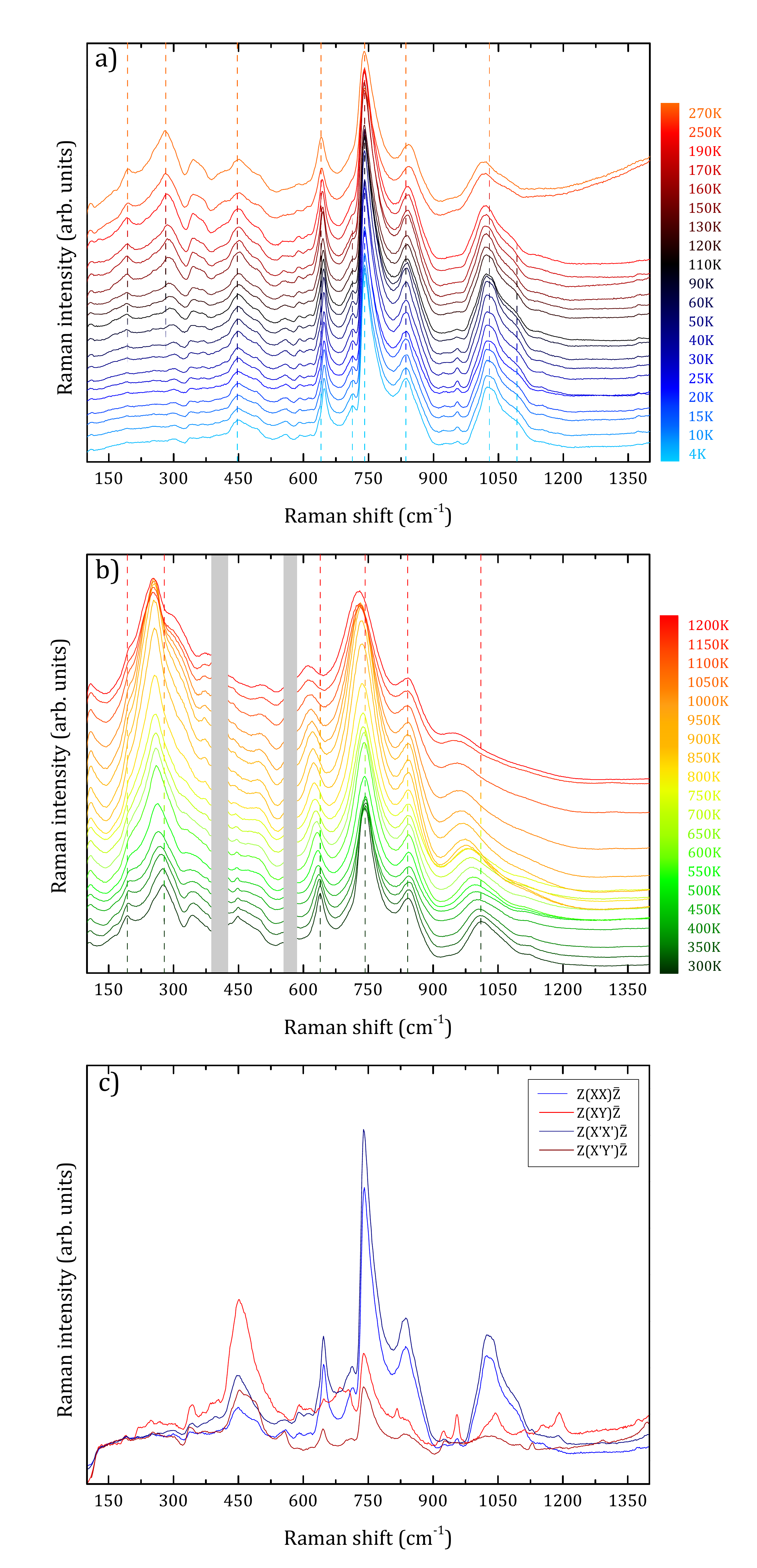}}
    \caption{Experimental Raman data showing a) the low-temperature and b) high-temperature dependences of the Raman spectra took in parallel polarization configuration (the crossed polarization spectra are shown in the Supplementary Material), the dotted lines are guidelines to follow the higher intensity modes that are discussed further on, and c) the polarized Raman measurements at 4~K, in the 4 polarization configurations described in the text.}
    \label{Raman_spectra}
\end{figure}

The polarized Raman spectra of our BaZrO$_3$ single crystals taken from 4~K to room temperature and up to 1200~K are shown on Figures~\ref{Raman_spectra}.a and \ref{Raman_spectra}.b respectively. 

The first result stemming from these spectra is that the behaviour of the spectra between 4 and 1200~K shows no evidence of a structural transition occurring in BaZrO$_3$ in that range of temperatures, indicating a very stable overall structure. 

At low temperature, approximately below 100~K, a large number of weak thin peaks appear, especially at high frequencies. Those peaks do not appear in our complementary measurements performed with a different excitation wavelength (532~nm), and we therefore conclude that they are not Raman peaks but rather originate from some luminescence processes. Their detailed assignment is beyond the scope of this work, but we note that they are very similar in positions, shape, and temperature behaviour to the lines observed in the Raman spectra of LaAlO$_3$, which have been assigned to various intrinsic defects~\cite{Saha2016}.  

While the spectrum consists mostly of broad features (reminiscent of second-order scattering), the main dominant peaks are quite sharp, and do not broaden significantly when the temperature increases: their full-width at half maximum (FWHM) is around 20-40~cm$^{-1}$ at 4~K and 40-60 at room temperature (Fig.\ref{Raman_spectra}.a), which is comparable to first-order phonon modes. Surprisingly, the frequencies of these peaks are shifting much less with temperature than commonly observed for thermal softening of phonon modes. As an example, the three dominant peaks that are sharper and more intense situated at 648, 741 and 838~cm$^{-1}$ at 4~K are found at 1200~K at 613, 728 and 843~cm$^{-1}$ respectively, corresponding to relative shifts of -4.5, -1.5 and +0.5 $\cdot 10^{-5}$.K$^{-1}$.

In order to separate the spectrum into its different symmetry components, we performed polarized Raman measurements. Choosing $x$, $y$ and $z$ along the cubic crystallographic axes, and labelling $x'$ and $y'$ axes rotated by 45$^\circ$ along the $z$ axis, the polarization components can be accessed in the following backscattering geometries (in Porto notation): 
\begin{eqnarray}
z(xx)z   & \rightarrow & \mathrm{A}_\mathrm{1g}+E_\mathrm{g}\nonumber\\ 
z(xy)z   & \rightarrow & \mathrm{T}_\mathrm{2g}\nonumber\\ 
z(x'x')z & \rightarrow & \mathrm{A}_\mathrm{1g}+T_\mathrm{2g}\nonumber\\
z(x'y')z & \rightarrow & \mathrm{E}_\mathrm{g}\nonumber
\end{eqnarray}
The spectra recorded in those 4 configurations at 4~K are shown in figure~\ref{Raman_spectra}.c. The E$_\mathrm{g}$ and T$_\mathrm{2g}$ spectra are found to be weaker. The broad peak at 450~$cm^{-1}$ appears more strongly in the T$_\mathrm{2g}$ spectrum, but no other feature can be conclusively assigned to these symmetries only. As a result, we will consider in the following that the spectrum has essentially the full symmetry A$_\mathrm{1g}$.

\section{Discussion: interpretation/origin of the Raman spectrum}

\subsection{High-frequency part of the spectrum}

\begin{figure}[htpb]
\resizebox{8.3cm}{!}{\includegraphics{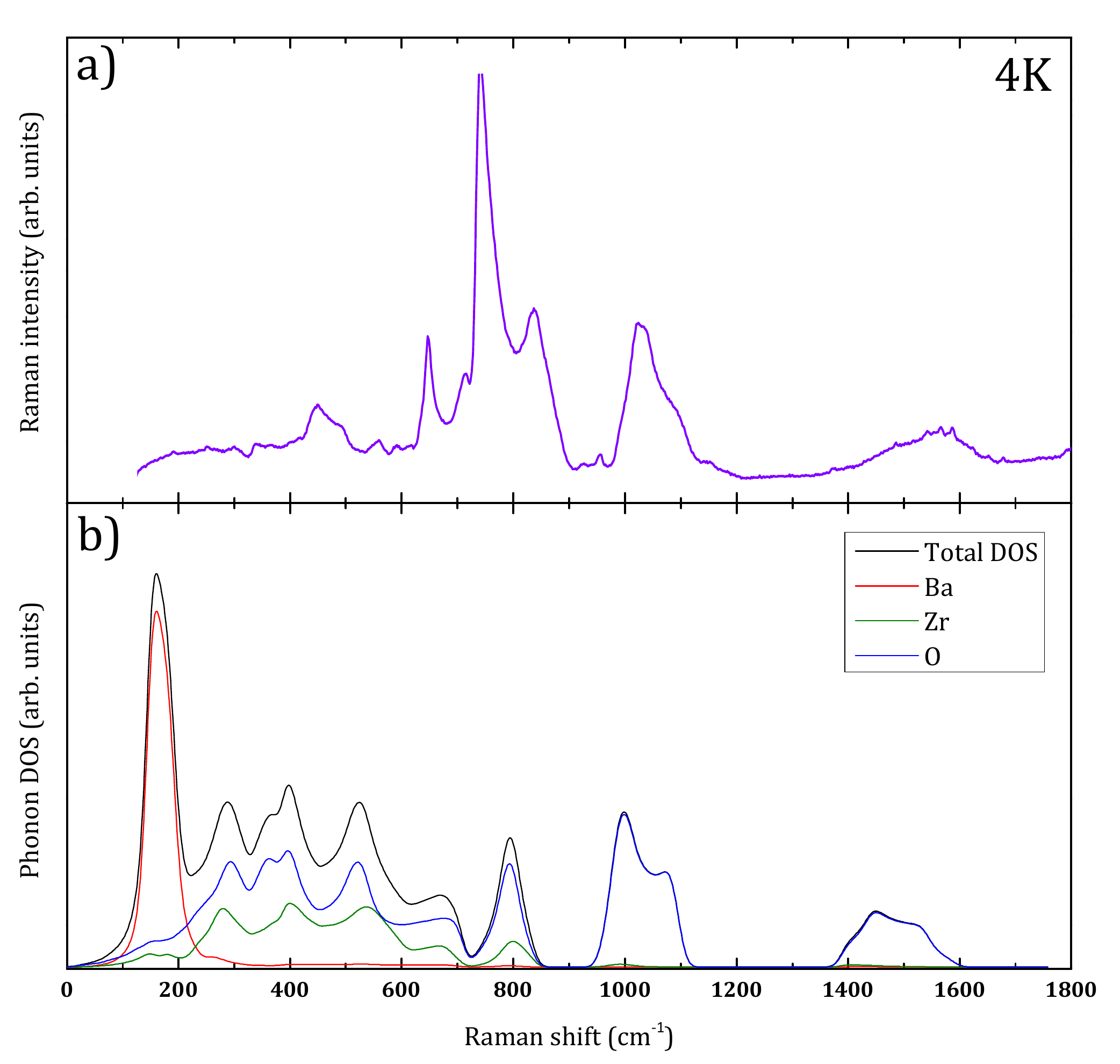}}
\caption{a) Raman spectrum of BaZrO$_3$ at 4~K, in the z(xx)z geometry.  b) Overtone density of states, whereby the respective contributions of different atoms are separated and shown alongside the total density of states.}
\label{2DOS}
\end{figure}

To discuss the interpretation of the Raman spectrum of BaZrO$_3$ and try to understand its origin, we would like to first focus on the high-frequency part of the spectrum (above 800~cm$^{-1}$). This range is above the (one-)phonon frequency range, so that the assignment of those features to second-order scattering with the creation (or annihilation) of two phonons is in fact the only possible. Since overtones are always Raman-allowed, we start by comparing in Figure~\ref{2DOS} the overtone total density of states with the experimental spectra taken at 4~K. The agreement is very good, and the experimental spectrum looks like a faithful image of the full doubled density of states. The phonon modes involved are the high energy modes corresponding to the octahedra stretching branches. There is only a small shift in frequency between the experimental spectrum and the total DOS, easily explained by the usual mismatch in volume yielded by DFT calculations.

Assignment of the kinks in the DOS/experimental spectrum can be done based on the shape of the spectrum. Of interest is the main peak at 1028~cm$^{-1}$ which can be assigned to the overtone of the IR-active TO mode and the 1090~cm$^{-1}$ to an overtone of the $R_3^+$ mode. The second kink around 1500~cm$^{-1}$ corresponds to the oxygen octahedra breathing mode, in the highest branch. Interestingly, we can notice that zone-boundary points account for the kinks in the spectrum, but do not give rise to increased intensity. Also, the $R$ point, from which the structural instability arises, does not play any particular role in the understanding of the high-frequency part of the Raman spectrum. 

\subsection{Low-frequency spectrum}

For the lower part of the spectrum (below 800~cm$^{-1}$), the overtone density of states clearly does not match the observed spectrum. In particular, it does not show any obvious assignment for the three dominant features of the experimental spectrum at 451, 648 and 741~cm$^{-1}$, and conversely, the experimental spectrum does not show any sign of the intense peak in the density of states associated to Ba motion. We discuss their possible origins below. 

Let's first consider the hypothesis of first-order scattering allowed by a local symmetry breaking, with the idea that the structure at the local scale is close to the tilted phases described previously. Within that model, we expect the most intense features of the Raman spectrum to correspond to frequencies of phonon modes that become Raman active in the tilted phases. However, the comparison of the frequencies of the phonon modes computed by DFT in the 3 possible distorted phase given in Table \ref{DFT_modes_freq} does not support this hypothesis. The three most intense experimental Raman modes at 741, 648 and 451~cm$^{-1}$ are not found in the DFT modes frequencies of any of the structural phases proposed. Therefore, the scenario involving first order Raman signal stemming from nano-domains of one of the three possible distorted structures does not provide a convincing explanation of our experimental phonon modes, unless one assumes other type of distortions.

Another hypothesis is to assign the peaks to the polar modes that are activated by disorder. This was for example proposed for the intense modes in BaSnO$_3$ that bears similarities with the BaZrO$_3$ peaks~\cite{stanislavchuk_electronic_2012}. We investigated that hypothesis by taking the IR experimental values of the TO and LO frequencies, extracted from refs. \onlinecite{nuzhnyy_broadband_2012, bennett_effect_2006} and comparing them to our experimental data. The frequencies do not match our data either (cf Table. III in the Suppl.). We therefore discard this hypothesis. More generally, we observe that there is a large gap in the phonon density of states between 553 and 692~cm$^{-1}$, so that it is nearly impossible to assign the peak at 648~cm$^{-1}$ to any first-order phonon scattering process. 

We therefore conclude that the low-frequency spectrum is also dominantly of second-order character. To assign the spectrum features more precisely to particular phonon modes, we kept the modes combinations with the closest frequencies to our experimental frequencies. These assignments are shown in Table \ref{Modes_assignments_cubic}. Interestingly, the two dominant peaks at 741 and 648~cm$^{-1}$ are assigned to polar modes at the zone center. In some cases, more than one combination can match the observed frequency. It must also be pointed out that assignment at low frequencies must be taken with care for two reasons. First of all, the real frequency of the soft mode in the cubic phase remains unknown. Considering its energy in the stable tilted phase (Table~\ref{DFT_modes_freq}), we can reasonably hypothesize that it is below 100~cm$^{-1}$, and unlikely to affect our interpretation of the high frequency spectrum. Secondly, the low frequency part may contain subtraction combinations that we are not considering here.

In addition, our assignments are based on the concordance between experimental and theoretical frequencies, which arguably leads to uncertainties in the mode attributions. In ref. \onlinecite{nuzhnyy_broadband_2012}, the discrepancy between experimental infrared and theoretical frequencies for the polar modes can go up to 30~cm$^{-1}$. In this work, the misfit between the experimental and theoretical maxima of the intense band around 1000~cm$^{-1}$ is just over 20~cm$^{-1}$, yielding an uncertainty of around 10~cm$^{-1}$ for the phonon modes frequencies calculated by DFT. Because of that, we decided to take a error margin of about 20~cm$^{-1}$ when searching for theoretical modes matching our Raman data.

The general temperature dependence of the spectrum is not conclusive in itself, but is consistent with a second-order assignment. The temperature dependence of Raman intensity can in principle allows to distinguish first-order from second-order scattering, because both have different dependence of the Bose-Einstein occupation factor $n$: $(1+n)$ for first-order, $(1+n)(2+n)$ for overtones and $(n+1)^2$ for combinations of different phonons~\cite{cardona_light_1982}. Qualitatively, this leads to an enhanced intensity at low frequencies and high temperatures, which is in agreement with what is observed in Fig.~\ref{Raman_spectra}.b. On the other hand, the Raman intensity can be expected to decrease with temperature, due to thermal expansion~\cite{Lucazeau2003}, with an unknown rate, which makes the analysis difficult in practice.
The temperature dependence of the peak positions is found to be equally inconclusive: they shows relative shifts of -4.5 and -1.5~$\cdot 10^{-5}$cm$^{-1}$.K$^{-1}$ respectively for the main peaks. Taking a literature value for the volumic thermal expansion of BaZrO$_3$: 2.1$\cdot 10^{-5}$cm$^{-1}$.K$^{-1}$~\cite{Zhao1991,Yamanaka2003}, we would obtain Grüneisen parameters of 2.1 and 0.7 respectively which can support both the first and second-order nature of the peaks.

Finally, we note that the pressure dependence of the Raman modes can be also ambiguous to distinguish between first-order from second-order processes. With increasing pressure, the room-temperature Raman spectrum of BaZrO$_3$ tends to vanish, which was taken as a support for a scenario where locally distortions give space to macroscopically cubic phase~\cite{chemarin_high-pressure_2000}. However, a classical second-order spectrum is also expected to decrease in intensity with pressure~\cite{Trallero-Giner2010} - as it was observed in SrTiO$_3$~\cite{Guennou2010} and CaTiO$_3$~\cite{Guennou2010a}, so that this observation alone is inconclusive.

\begin{table*}[th!]
	\centering
	\caption{Experimental frequencies of the intense Raman peaks and their possible attribution to second order phonon modes combinations calculated by DFT. The comparison with the experimental infrared frequencies derives from \onlinecite{nuzhnyy_broadband_2012} are also shown. The Raman frequencies are given at 4~K, unless specified otherwise. Overtones are explicitly indicated while $\oplus$ indicates a combination of different phonon branches.}
	\label{Modes_assignments_cubic}
	\setlength{\extrarowheight}{3pt} 
	\begin{tabular}{c|c|c|l|c}
		\hline\hline
		Experiment & \multicolumn{3}{l|}{DFT theoretical assignment - second order} & Literature (exp)\\
		\hline
		freq. (cm$^{-1}$) & freq. (cm$^{-1}$) & Combination & Associated atomic displacement & Nuzhnyy et al.\cite{nuzhnyy_broadband_2012}\\
		\hline
		451 & 451 & R$_5^+\oplus\mathrm R_5^+$ & Ba (antiparallel) + octahedra rotations & \\
		\hline
		648 & 636  & $\Gamma_4^-\oplus\Gamma_4^-$ & Last(LO) + Axe(TO) zone center modes & 661 \\
		\hline
		713 & 707  & overtone R$_5^+$ & ``scissoring" mode of octahedra & \\
		\hline
		\multirow{3}{*}{741} & 732 & overtone $\Gamma_4^-$ & Slater(LO) zone center mode & 760\\
		 & 745 & X$_5^-\oplus\mathrm X_5^-$ & in-plane antiparallel displacement of apical O & \\
		  & & & + antiparallel apical oxygen displacements & \\
		\hline
		838 & 840 & overtone X$_3^-$ & displacements of Zr against in-plane oxygens & \\
		 & & & Last (LO) + Axe (LO) & 833\\
		\hline
		1028 & 1006 & overtone $\Gamma_4^-$ & Zone center Axe(TO) mode & 1040\\
		1045 & 1059 & $\Gamma_4^-\oplus\Gamma_4^-$ & Zone center Axe(LO) + Slater(LO) modes & 1072\\
		1090 & 1083 & overtone R$_3^+$ & Jahn-Teller-like octahedra distortions & \\
		\hline
		\multirow{2}{*}{1573} & 1538 & overtone M$_1^+$ & oxygen octahedra breathing mode (x,y)\\
		& 1597 & overtone R$_1^+$ &  oxygen octahedra breathing mode\\
		\hline
		\multirow{2}{*}{193 (Room T)} & 193 & overtone $\Gamma_4^-$ & Last(TO) zone center mode & 232\\
		& 195 & overtone R$_5^+$ & antiparallel Ba displacements & \\
		\hline 
		\multirow{3}{*}{277 (Room T)} & 277 & $\Gamma_4^-\oplus\Gamma_5^-$ & Last(TO) + shear mode octahedra & \\
		& 277 & X$_5^+\oplus\mathrm X_5^+$ & antiparallel Ba displacements (x,y) & \\
		& & & + antiparallel oxygens (x,y) displacements & \\
		& 280 & overtone M$_3^-$ & antiparallel Zr displacements (z) & \\
		\hline
		\multirow{2}{*}{253 (1200~K)} & 266 & overtone $\Gamma_4^-$ & Last(LO) zone center mode & 282\\
		& 262 & overtone X$_1^+$ & antiparallel Ba displacements (x) & \\
		\hline\hline
	\end{tabular}
\end{table*}

\subsection{Relation with local disorder}

In summary, we find that the whole Raman spectrum of BaZrO$_3$ can be explained by second-order Raman scattering. We do not find that the phonons at the $R$-point, where BaZrO$_3$ has its structural instability, play a special role. In particular, we find no evidence for scattering directly related to octahedra tilts, and no direct support for the hypothesis of distorted nano-domains. 

On the other hand, it is still very likely that the overall intensity of the scattering finds its origin in another type of local disorder. Let us first point out that the presence of a strong Raman spectrum in cubic perovskites is not a general property. This can be appreciated by comparing the relative strengths of second vs. first order spectra: typical perovskites such as LaAlO$_3$ or PbTiO$_3$ exhibit in their cubic phase a Raman spectrum that is extremely weak compared to the intensity of the first-order Raman modes allowed in their low-symmetry phase. In contrast, in SrTiO$_3$ or CaTiO$_3$, both contribution are or equal strength. Those compounds also share common experimental signatures related to local disorder of the Ti$^{4+}$ ion in the octahedral cage, with the presence of intense diffuse scattering~\cite{Kopecky2019}, and distinctive features in XANES~\cite{Itie2006}. We therefore expect that studies of the local structure in BaZrO$_3$ will be insightful to reveal the details and dynamics and disorder, and we note that a previous study has already proposed anomalous Ba displacements, revealed by anomalously large Debye–Waller (DW) factor in BaZrO$_3$ structural refinements~\cite{lebedev_structural_2013}.  

\section{Conclusion}

In summary, we have provided measurements of the Raman spectrum of BaZrO$_3$ single crystals in a wide temperature range, and analyzed it with the light of comprehensive lattice dynamical calculations of BaZrO$_3$ in its cubic phase as well as in its distorted tilted phases. We clearly identified a major contribution from second-order Raman scattering, especially overtones of the octahedra breathing modes at high frequencies. Other salient features of the Raman spectrum have also been assigned to overtones and mode combinations. Raman spectroscopy therefore does not directly support the hypothesis of locally tilted nano-domains, although local disorder is still believed to play a role in the general intensity of the Raman scattering. In addition, we have attracted attention to the expected similarities between the Raman spectrum of all tilted phases, which calls for some care in the claims for transition to orthorhombic phases in high-pressure measurements of BaZrO$_3$ and SrTiO$_3$. Furthermore, the low-frequency lattice dynamics at the zone-boundary and its behaviour with temperature remains to be experimentally determined and will be the object of further work.

\begin{acknowledgments}
This work was supported by the Innovative Training Networks (ITN) – Marie Skłodowska-Curie Actions-European Joint Doctorate in Functional Material Research (EJD-FunMat) project (no. 641640). DFT-based calculations have been performed on the NIC4 cluster hosted at the University of Li\`ege, within the `Consortium des \'Equipements de Calcul Intensif' (C\'ECI), funded by F.R.S-FNRS (Grant No 2.5020.1) and by the Walloon Region. C.T., M.G., J.K. acknowledge financial support from the Fond National de Recherche Luxembourg through a
PEARL grant (FNR/P12/4853155/Kreisel). The work at the University of Warwick was supported by the EPSRC, UK, Grant EP/M028771/1.
\end{acknowledgments}

%
\end{document}